\documentclass[twocolumn,aps,showpacs]{revtex4}
\usepackage{amssymb}
\usepackage{epsf}

\usepackage{amsmath}

\def\lsim{\lower -0.3ex \hbox{$<$} \kern -0.75em \lower 0.7ex \hbox{$\sim$}}
\def\gsim{\lower -0.3ex \hbox{$>$} \kern -0.75em \lower 0.7ex \hbox{$\sim$}}

\def\Vec#1{{\bf #1}}

\def\vare{\varepsilon}

\def\partd#1#2{\frac{\partial #1}{\partial #2}}

\begin{document}

\title{Gate-induced interlayer asymmetry in ABA-stacked trilayer graphene}
\author{Mikito Koshino$^{1}$ and Edward McCann$^{2}$}
\affiliation{
$^{1}$Department of Physics, Tokyo Institute of
Technology, 2-12-1 Ookayama, Meguro-ku, Tokyo 152-8551, Japan\\
$^{2}$Department of Physics, Lancaster University, Lancaster, LA1
4YB, UK}

\begin{abstract}
We calculate the electronic band structure of ABA-stacked trilayer
graphene in the presence of external gates, using a
self-consistent Hartree approximation to take account of
screening. In the absence of a gate potential, there are separate
pairs of linear and parabolic bands at low energy. A gate field
perpendicular to the layers breaks mirror reflection symmetry with
respect to the central layer and hybridizes the linear and
parabolic low-energy bands,
leaving a chiral Hamiltonian essentially
different from that of monolayer or bilayer graphene.
%leaving, for large gate fields, just
%two bands in the vicinity of zero energy, which support chiral
%quasiparticles.
Using the self-consistent Born approximation, we
find that the density of states and the minimal conductivity in
the presence of disorder generally increase as the gate field
increases, in sharp contrast with bilayer graphene.
\end{abstract}


\maketitle

\begin{figure}[t]
%\centerline{\epsfxsize=1.0\hsize \epsffile{figure1.eps}}
%\centerline{\epsfxsize=0.95\hsize \epsffile{fig1.eps}}
\centerline{\epsfxsize=0.95\hsize \epsffile{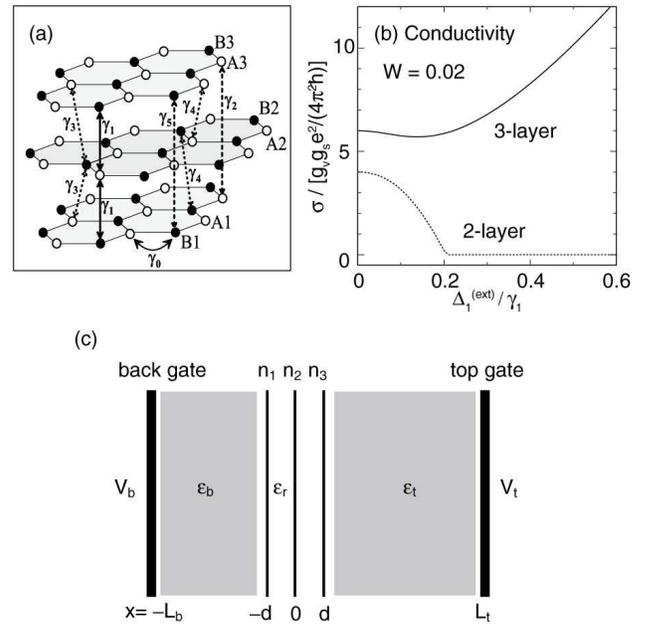}}
\caption{(a) Schematic of the ABA-stacked trilayer lattice
containing six sites in the unit cell, $A$ (white circles) and $B$
(black circles) on each layer, showing the
Slonczewski-Weiss-McClure parameterization \cite{dressel02} of
relevant couplings $\gamma_0$ to $\gamma_5$. (b) The conductivity
versus external asymmetry $\Delta_1^{\rm (ext)}$, calculated for
trilayer and bilayer graphene using the self-consistent Born
approximation and the band model including $\gamma_0$ and
$\gamma_1$. (c) Schematic of trilayer graphene (three thin black
lines at $x = -d$, $0$, $d$) with top and bottom gates (thick
black lines at $x= L_t, - L_b$) separated from the trilayer by
dielectric media (gray shaded areas). } \label{fig:1}
\end{figure}

Pioneering experiments \cite{novo04,novo05,zhang05,novo06}
demonstrated graphene-based transistors using a back gate to vary
the carrier density continuously from electron to hole channels,
with a minimal conductivity for nominally-zero carrier density.
The switching of a graphene-based transistor would be improved by
opening an energy gap between the conduction and valence bands,
possibly by lateral confinement of electrons in etched structures
\cite{han07,miao07,stamp08,pono08} or by employing gates to induce
interlayer asymmetry in bilayer graphene
\cite{mcc06a,lu06,guinea06,ohta06,mcc06b,min07,castro07,oost08}.
Recently, experimental attention has turned towards the properties
of ABA-stacked trilayer graphene, Fig.~\ref{fig:1}(a),
\cite{ohta07,guett08,crac08}. Theory suggests that the bands are
of two separate types
\cite{lu06,latil06,part06,guinea06,Kosh_mlg,aoki07}: two
almost-linear bands reminiscent of the bands in monolayer graphene
and four parabolic bands similar to those in bilayer graphene.
This raises the expectation that the electronic behavior will
display no new features as compared to monolayer or bilayer
graphene.

In this paper, we show theoretically that the response of
ABA-stacked trilayer graphene to external gate potentials is in
fact qualitatively different from that in mono- or bi-layer
graphene. We use an effective-mass model to self-consistently
determine the electronic band structure and we show how the
breaking of mirror reflection symmetry by interlayer asymmetry
causes hybridization of the linear and parabolic bands. Rather
than opening a gap, as in bilayer graphene \cite{mcc06a}, this
leaves two bands near zero energy which support chiral
quasiparticles. Employing a self-consistent Born approximation to
estimate the minimal conductivity as a function of interlayer
asymmetry, we find that the conductivity generally increases as
asymmetry increases, in sharp contrast with bilayer graphene as
illustrated in Fig.~\ref{fig:1}(b).

%%%%%%%%%%%%%%%%%%%%%%%%%%%%%%%%%%%%%%%%%%%%%%%%%%%%%%%%%%%%%%%%%%%%%%%%%%%%%%
%%%%%%%%%%%%%%%%%%%%%%%%%%%%%%%%%%%%%%%%%%%%%%%%%%%%%%%%%%%%%%%%%%%%%%%%%%%%%%

A description of trilayer graphene in the presence of external
gates must include {\em two} parameters that take into account
differences in the potentials $V_1$, $V_2$, and $V_3$ of the three
layers. The first, $\Delta_1 = -e (V_1 - V_3)/2$, describes the
average energy difference between each adjacent layer
\cite{lu06,guinea06,aoki07}, while the second, $\Delta_2 = -e (V_1
- 2V_2 + V_3)/6$, describes the difference between the energy of
the central layer and the average of the outer layers. We model
the effect of back and top gates by considering the trilayer as
three conducting parallel plates as illustrated in
Fig.~\ref{fig:1}(c), with respective electron densities $n_1, n_2$
and $n_3$, located at $x = -d$, $0$, and $+d$, respectively, where $d$
is the interlayer spacing, and the permittivity of the trilayer
interlayer spaces (without the screening effect of $\pi$-band
electrons of the trilayer graphene) is $\varepsilon_r$. The back
(top) gate at $x = - L_b$ ($x= + L_t$), held at potential $V_b$
($V_t$), is separated from the trilayer by a dielectric medium
with relative permittivity $\varepsilon_b$ ($\varepsilon_t$).
Using elementary electrostatics, we relate the external gate
potentials, the electron densities on the layers, and the
interlayer asymmetry parameters:
\begin{eqnarray}
&&
\frac{\varepsilon_b V_b}{L_b}
+ \frac{\varepsilon_t V_t}{L_t} = e \left( n_1 + n_2 + n_3 \right),
 \label{eqa} \\
&& \Delta_1 =
\frac{\varepsilon_t V_t}{L_t}- \frac{\varepsilon_b V_b}{L_b}
+ \frac{e^2d}{2\varepsilon_r}(n_1-n_3),
%+ \frac{e^2c_0}{4\varepsilon_r}(n_1-n_3),
\label{eqb} \\
&& \Delta_2 = - \frac{e^2d}{6\varepsilon_r}n_2 \, .
%&& \Delta_2 = - \frac{e^2c_0}{12\varepsilon_r}n_2 \, .
\label{eqc}
\end{eqnarray}
In the following, we use the total electron density $n_{\rm
tot} = n_1+n_2+n_3$ and $\Delta_1^{\rm (ext)}$ as external
parameters instead of $V_t$ and $V_g$, where
$\Delta_1^{\rm (ext)} = \varepsilon_t V_t/L_t-
\varepsilon_b V_b/L_b$ is the value of $\Delta_1$ that would occur
if screening were negligible.

We model ABA-stacked trilayer graphene as three coupled honeycomb
lattices including pairs of inequivalent sites $\{ A1 , B1 \}$,
$\{ A2 , B2 \}$, and $\{ A3 , B3 \}$ in the bottom, center, and
top layers, respectively. The layers are arranged according to
Bernal ($A$-$B$) stacking \cite{dressel02}, Fig.~\ref{fig:1}(a),
such that sites $B1$, $A2$, and $B3$ lie directly above or below
each other. We employ an effective-mass model adopting the
Slonczewski-Weiss-McClure parameterization, \cite{dressel02} where
each parameter is related to relevant coupling in the
tight-binding model: $\gamma_0$ describes nearest-neighbor
($Ai$-$Bi$ for $i= \{1,2,3\}$) coupling within each layer,
$\gamma_1$ describes strong nearest-layer coupling between sites
($B1$-$A2$ and $A2$-$B3$) that lie directly above or below each
other, $\gamma_3$ ($\gamma_4$) describes weaker nearest-layer
coupling between sites $A1$-$B2$ and $B2$-$A3$ ($A1$-$A2$,
$B1$-$B2$, $A2$-$A3$, and $B2$-$B3$). With only these couplings,
there would be a degeneracy point at each of two inequivalent
corners, $K_{\pm}$, of the hexagonal Brillouin zone \cite{kpoints}
but this degeneracy is broken by next-nearest-layer coupling
$\gamma_2$ (between $A1$ and $A3$), $\gamma_5$ (between $B1$ and
$B3$) and $\delta$, which is the on-site energy difference between
$A1,B2,A3$ and $B1,A2,B3$. Note that the parameter $\Delta$ often
used in models of three-dimensional (3D) 
graphite is given by $\Delta = \delta +
\gamma_2 - \gamma_5$.
%$\delta = \Delta - \gamma_2 + \gamma_5$.
In trilayer graphene, the presence of a surface may induce a
modification in the value of the band parameters as compared to
those in bulk graphite. Here, parameter $\Delta_2= -e (V_1 - 2V_2
+ V_3)/6$, takes into account a possible difference between the
energy of the central layer and the average of the outer layers,
and, in general, surface effects may contribute to a non-zero
value of $\Delta_2$.

In a basis with components $\psi_{A1}$, $\psi_{B1}$, $\psi_{A2}$ ,
$\psi_{B2}$, $\psi_{A3}$, $\psi_{B3}$, the ABA-stacked trilayer
Hamiltonian is
\begin{eqnarray}
{\widetilde H} = \left(%
%\begin{array}{cccccc}
%  U_1 & v \pi^{\dag} & v_4 \pi^{\dag} & v_3 \pi & \gamma_2 & 0 \\
%  v \pi & U_1 & \gamma_1 & v_4 \pi^{\dag} & 0 & \gamma_5 \\
%  v_4 \pi & \gamma_1 & U_2  & v \pi^{\dag} & v_4 \pi & \gamma_1 \\
%  v_3 \pi^{\dag} & v_4 \pi & v \pi & U_2 & v_3 \pi^{\dag} & v_4 \pi \\
%  \gamma_2 & 0 & v_4 \pi^{\dag} & v_3 \pi & U_3 & v \pi^{\dag} \\
%  0 & \gamma_5 & \gamma_1 & v_4 \pi^{\dag} & v \pi & U_3 \\
%\end{array}%
\begin{array}{cccccc}
  U_1 & v \pi^{\dag} & -v_4 \pi^{\dag} & v_3 \pi & \gamma_2/2 & 0 \\
  v \pi & U_1 + \delta & \gamma_1 & -v_4 \pi^{\dag} & 0 & \gamma_5/2 \\
  -v_4 \pi & \gamma_1 & U_2 + \delta & v \pi^{\dag} & -v_4 \pi & \gamma_1 \\
  v_3 \pi^{\dag} & -v_4 \pi & v \pi & U_2 & v_3 \pi^{\dag} & -v_4 \pi \\
  \gamma_2/2 & 0 & -v_4 \pi^{\dag} & v_3 \pi & U_3 & v \pi^{\dag} \\
  0 & \gamma_5/2 & \gamma_1 & -v_4 \pi^{\dag} & v \pi & U_3 + \delta \\
\end{array}%
\right) , \label{h1}
\end{eqnarray}
where operator $\pi = \xi p_x + i p_y$
%and $\pi^{\dag} = \xi p_x - i p_y$
is related to the in-plane momentum ${\bf p} = ( p_x ,p_y )$
%measured with respect to the $K$ point
\cite{kpoints},
effective velocities are $v= ( \sqrt{3}/2 ) a\gamma_{0}/\hbar$,
$v_3 = ( \sqrt{3}/2 ) a\gamma_{3}/\hbar$, and $v_4 = ( \sqrt{3}/2
) a\gamma_{4}/\hbar$, $U_i = -e V_i$, and $\xi = \pm 1$ is the
valley index $K_\pm$.
Exploiting mirror reflection symmetry of the lattice
in the plane of its central layer, Fig.~\ref{fig:1}(a), we perform
a unitary transformation to a basis consisting of linear
combinations of the atomic orbitals \cite{Kosh_mlg}, namely
$[\psi_{A1} - \psi_{A3}]/\sqrt{2}$, $[\psi_{B1} -
\psi_{B3}]/\sqrt{2}$, $[\psi_{A1} + \psi_{A3}]/\sqrt{2}$,
$\psi_{B2}$, $\psi_{A2}$, $[\psi_{B1} + \psi_{B3}]/\sqrt{2}$:
\begin{eqnarray}
&& \hspace{-4mm}
H = \left(%
\begin{array}{cc}
  H_m & D \\
  D^T & H_b \\
\end{array}%
\right) ,
\quad
D = \left(%
\begin{array}{cccc}
\Delta_1 & 0 & 0 & 0 \\
0 & 0 & 0 & \Delta_1 \\
\end{array}%
\right) ,
\label{h2} \\
&& \hspace{-4mm}
H_m = \left(%
\begin{array}{cc}
  \Delta_2 - \gamma_2/2 & v \pi^{\dag} \\
  v \pi & \Delta_2 - \gamma_5/2 + \delta \\
\end{array}%
%\begin{array}{cc}
%  \Delta_2 - \gamma_2 & v \pi^{\dag} \\
%  v \pi & \Delta_2 - \gamma_5 \\
%\end{array}%
\right) ,
\label{hm} \\
&& \hspace{-4mm}
H_b = \left(%
\begin{array}{cccc}
\Delta_2 + \gamma_2/2 & \sqrt{2} v_3 \pi & -\sqrt{2} v_4 \pi^{\dag} & v \pi^{\dag} \\
\sqrt{2} v_3 \pi^{\dag} & - 2 \Delta_2 & v \pi & -\sqrt{2} v_4
\pi \\
-\sqrt{2} v_4 \pi & v \pi^{\dag} & - 2 \Delta_2 + \delta & \sqrt{2} \gamma_1 \\
v \pi & -\sqrt{2} v_4 \pi^{\dag} & \sqrt{2} \gamma_1 &
\Delta_2 + \gamma_5/2 + \delta\\
\end{array}%
%\begin{array}{cccc}
%\Delta_2 + \gamma_2 & \sqrt{2} v_3 \pi & \sqrt{2} v_4 \pi^{\dag} & v \pi^{\dag} \\
%\sqrt{2} v_3 \pi^{\dag} & - 2 \Delta_2 & v \pi & \sqrt{2} v_4
%\pi \\
%\sqrt{2} v_4 \pi & v \pi^{\dag} & - 2 \Delta_2 & \sqrt{2} \gamma_1 \\
%v \pi & \sqrt{2} v_4 \pi^{\dag} & \sqrt{2} \gamma_1 & \Delta_2 + \gamma_5 \\
%\end{array}%
\right) , \label{hb}
\end{eqnarray}
where the average on-site energy $[U_1 + U_2 + U_3]/3$ has been
set equal to zero. The Hamiltonian $H$ has a $2 \times 2$ block
$H_m$ and a $4 \times 4$ block $H_b$ on the diagonal, connected by
a simple off-diagonal block $D$. Block $H_m$ is
similar to the Dirac-type Hamiltonian of monolayer graphene and it
contributes two bands near zero energy whereas
block $H_b$ is reminiscent of the Hamiltonian of bilayer graphene
\cite{mcc06a}, except that terms proportional to $\gamma_{1}$,
$\gamma_{3}$, and $\gamma_{4}$ appear with a factor $\sqrt{2}$
\cite{Kosh_mlg}. The latter gives two bands split away from zero
by energy $\pm \sqrt{2} \gamma_1$ and two bands near zero energy.

The monolayer-like block has wave functions possessing odd mirror
reflection symmetry, while the wave functions of the bilayer part
are even. Since the interlayer asymmetry $\Delta_1$ is the only
parameter that breaks mirror reflection symmetry, its role is
qualitatively different from the other parameters, coupling the
monolayer-like and bilayer-like blocks. For large $\Delta_1$, two
of the low-energy bands, related to orbitals $[\psi_{A1} -
\psi_{A3}]/\sqrt{2}$ and $[\psi_{A1} + \psi_{A3}]/\sqrt{2}$, split
away from zero by energy $\epsilon \approx \pm \Delta_1$ at the
$K$ point, leaving only two bands near zero,
associated with
$\Psi^{\prime} = \left( [\psi_{B1} - \psi_{B3}]/\sqrt{2} , \psi_{B2} \right)^{T}$.
%$[\psi_{B1} - \psi_{B3}]/\sqrt{2}$ and $\psi_{B2}$.
To obtain an approximate Hamiltonian $H_{\rm eff}$ for $\Psi'$,
we denote $H_2$ as the diagonal block of Hamiltonian $H$ corresponding to
these two low-energy components,
$H_4$ as the $4 \times 4$ diagonal block corresponding to
the high-energy components, and $V$ as the off-diagonal $2\times 4$ block
coupling $H_2$ and $H_4$.
The Schr\"odinger equation for $\Psi^{\prime}$
can be expanded up to first order in $\vare$ as
$[H_2 - V H_4^{-1}V^\dagger] \Psi^{\prime} = \vare S \Psi^{\prime}$
with $S\equiv 1 + V H_4^{-2} V^\dagger$.
Then, the effective Hamiltonian for $\Psi = S^{1/2}\Psi^{\prime}$ becomes
${H}_{\rm eff} \approx S^{-1/2} [H_2 - V H_4^{-1}V^\dagger] S^{-1/2}$.
For the moment, we focus on the role of $\Delta_1$ by
considering $\Delta_2 = \gamma_2 = \gamma_{3} =
\gamma_{4} = \gamma_5 = \delta = 0$.
%%%%%%
For large enough $\Delta_1$
($|\gamma_1| \gg |\Delta_1| \gg  |\epsilon|$),
$H_{\rm eff}$ is written as
\begin{eqnarray*}
H_{\rm eff} &\approx&
\left(
\begin{array}{cc}
  0 & X^{\dagger} \\
  X & 0
\end{array}%
\right) \, , \\
X &=& - \frac{\Delta_1 v \pi}{\sqrt{2} \gamma_1}
\left( 1 - \frac{v^2\pi\pi^\dagger}{\Delta_1^2}\right)
\left( 1 + \frac{v^2\pi\pi^\dagger}{\Delta_1^2}\right)^{-1/2}\, .
%X &=& - \frac{\Delta_1 v \pi}{\sqrt{2} \gamma_1}\left( 1 - \frac{v^2\pi \pi^{\dag}}{\Delta_1^2}
%  \right) \left( 1 + \frac{v^2\pi \pi^{\dag}}{\Delta_1^2}\right)^{-1/2}\, .
\end{eqnarray*}
For plane-wave eigenstates at zero magnetic field,
$\pi \pi^\dagger$ is just a number, $p^2$. The first factor
of $\pi = \xi p_x + i p_y$ in operator $X$ ensures that
such eigenstates are chiral,
$\Psi = \left(
e^{-i\xi \theta/2} , \mp \xi e^{i\xi \theta/2} \right)^{T}/\sqrt{2}$
with $\theta = \tan^{-1}(p_y/p_x)$.
The expression for the eigenenergies is
\begin{eqnarray}
\epsilon \approx \pm \frac{vp}{\sqrt{2} \gamma_1} \frac{\left( v^2 p^2 -
\Delta_1^2 \right)}{\sqrt{v^2 p^2 + \Delta_1^2}} ,
\label{eq:overlap}
\end{eqnarray}
which generalizes Eq.~(22) of Ref.~\onlinecite{guinea06}, showing
that there is a small overlap $\delta \epsilon \sim \Delta_1^2 /
\gamma_1$ between the two low-energy bands that cross at $p =
\Delta_1 / v$ \cite{guinea06}. This behavior contrasts with that
of bilayer graphene, where interlayer asymmetry introduces an
energy gap between the low-energy bands \cite{mcc06a} and tends to
suppress the chiral nature of quasiparticles in them.

For given parameters $\Delta_1$, $\Delta_2$ and fixed total
density $n_{\rm tot}$, the electron densities $n_1,n_2$ and $n_3$
may be determined by summing $(|\psi_{Ai}|^2+|\psi_{Bi}|^2)/L^2$
over the occupied eigenstates of the Hamiltonian (\ref{h1}).
However, such densities are also related to $\Delta_1$ and
$\Delta_2$ through Eqs.~(\ref{eqb}) and (\ref{eqc}), so it is
necessary to solve this set of equations self-consistently in
order to obtain values of $\Delta_1$ and $\Delta_2$ for given
external parameters $n_{\rm tot}$  and $\Delta_1^{\rm (ext)}$. A
similar procedure has been applied to bilayer graphene
\cite{mcc06b,min07,castro07} and to many-layered graphene
\cite{guinea07}. This Hartree approximation neglects effects
including exchange interaction, possible deformation of atomic
orbitals in the applied electric field, and the role of $\sigma$
orbitals in screening, but comparison with density functional
theory \cite{min07} in bilayers suggests that it is qualitatively
accurate.

For $\gamma_2 = \gamma_{3} = \gamma_{4} = \gamma_5 = \delta = 0$
with $n_{\rm tot} = 0$, it is possible to perform a linear
response calculation for infinitely small $\Delta_1^{\rm (ext)}$.
Within the first order in $\Delta_1$, we have $n_1-n_3 = \Pi
\Delta_1$ with
\begin{eqnarray}
\Pi
=
\sum_{\alpha,\alpha'}
f(\vare_{\alpha})
\frac{2|
\langle \alpha |
\partd{H}{\Delta_1}
%(\partial H/\partial\Delta_1)
| \alpha'\rangle|^2}
{\vare_{\alpha}-\vare_{\alpha'}}
 = - \frac{g_vg_s}{2\sqrt{2}\pi}
\frac{\gamma_1}{(\hbar v)^2},
\label{eq_pi}
\end{eqnarray}
where $g_s=g_v=2$ are the spin and valley degeneracies, respectively,
$| \alpha\rangle$ and $\vare_{\alpha}$
are the eigenstates and eigenenergy of the Hamiltonian without
$\Delta_1$ or $\Delta_2$,
$f(\vare)$ is the Fermi distribution function with zero Fermi energy.
Using Eqs.~(\ref{eqb}) and (\ref{eqc}),
we obtain the self-consistent solution
$\Delta_1 = \Delta_1^{\rm (ext)}/\vare_{\rm eff}$
with $\vare_{\rm eff} = 1 - (e^2 d/2\vare_r)\Pi$.
$\Delta_2$ is never induced.
Typical parameters $v=1.0\times 10^6$ m/s,
$\gamma_1=0.4$ eV, $d=0.334$ nm, %$c_0=0.668$ nm,
$\vare_r = 2$ give $1/\vare_{\rm eff} \approx 0.61$.

%%%%%%%%%%%%%%%%%%%%%%%%%%%%%%%%%%%%%%%%%%%%%%%%%%%%%%%%%%%%%%%%%%%%%%%%%%%%%%
\begin{figure}[t]
%\centerline{\epsfxsize=0.9\hsize \epsffile{fig_band.eps}}
%\centerline{\epsfxsize=0.9\hsize \epsffile{fig2.eps}} 
\centerline{\epsfxsize=0.9\hsize \epsffile{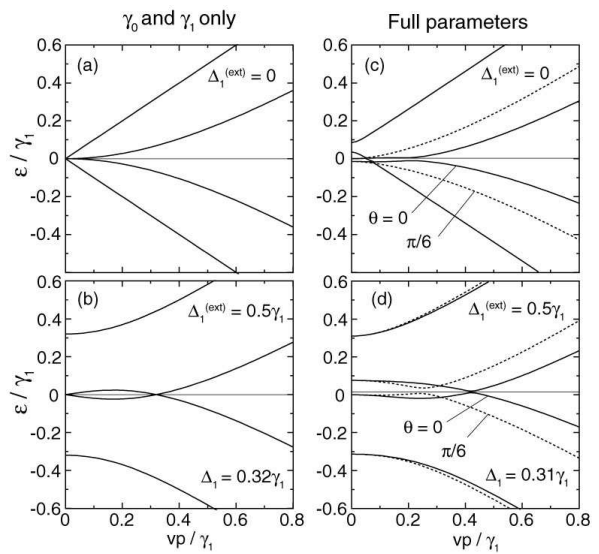}} 
\caption{
Self-consistently calculated band structures in trilayer graphene
near the $K$ point, with $n_{\rm tot}=0$. Left plots are for the
model including only $\gamma_0$ and $\gamma_1$, with (a) no
asymmetry $\Delta_1^{\rm (ext)} = 0$ and (b)  finite external
asymmetry $\Delta_1^{\rm (ext)} = 0.5\gamma_1$. Right plots are
for the full parameter model including $\gamma_i\,(i=1,2,\cdots5),
\delta$, with (c) $\Delta_1^{\rm (ext)} = 0$ and (d)
$0.5\gamma_1$. Dashed and solid curves represent $\theta =0$ and
$\pi/6$. The self-consistently calculated value of $\Delta_1$ %and $\Delta_2$
is shown in the lower side of each plot.
The thin horizontal line shows the Fermi energy.
}
\label{fig:2}
\end{figure}
%%%%%%%%%%%%%%%%%%%%%%%%%%%%%%%%%%%%%%%%%%%%%%%%%%%%%%%%%%%%%%%%%%%%%%%%%%%%%%

To determine the band structure
taking into account all the parameters, we find
$\Delta_1$ and $\Delta_2$ self-consistently
by employing an iterative numerical approach.
We first use $\Delta_1 = \Delta_1^{\rm (ext)}$
and $\Delta_2 = 0$ as initial values
in the Hamiltonian Eq.~(\ref{h1}) and
determine the Fermi energy so that the total density is
equal to $n_{\rm tot}$.
Then we calculate $n_i$ ($i=1,2,3$) from the occupied eigenstates,
which give a new set of $\Delta_1$ and $\Delta_2$ through
Eqs.~(\ref{eqb}) and (\ref{eqc}).
We iterate this process until $\Delta_1$ and $\Delta_2$
converge.

Figure~\ref{fig:2}
(a) and (c) show the self-consistent band structures at zero
external field $\Delta^{\rm (ext)} = 0$ and zero doping $n_{\rm
tot} = 0$. To illustrate the role of the extra band parameters we
compare (a) the simple model including only $\gamma_0, \gamma_1$
and (c) the full-parameter model with $\gamma_2 = -0.05\gamma_1$,
$\gamma_5 = 0.1\gamma_1$, $\delta = 0.125\gamma_1$, $v_3(\propto
\gamma_3) = 0.1 v$ and $v_4(\propto \gamma_4)  = 0.014 v$ (typical
values quoted for bulk graphite \cite{dressel02}). The plots show
the vicinity of zero energy, covering the monolayer-like band and
the lower branches of the bilayer-like band. In (c), we see that
$\gamma_2$, $\gamma_5$ and $\delta$ shift the center of the
monolayer-like band upward in energy relatively to the
bilayer-like band. Also, the trigonal warping effect due to
$\gamma_3$ is observed as a difference between
$\theta = \tan^{-1}(p_y/p_x) = 0$
and $\pi/6$ \cite{mcc06a}.
Figures~\ref{fig:2}(b) and
\ref{fig:2}(d) display the corresponding plots in the presence of a finite
external field $\Delta_1^{\rm (ext)} = 0.5\gamma_1$. The values of
$\Delta_1$ determined self-consistently are shown in the lower
side of each plot. In every case the screening ratio
$\Delta_1/\Delta_1^{\rm (ext)}$ is about 0.6, which is close to
the linear response theory.
For Fig. \ref{fig:2}  (b), where only
$\gamma_0$ and $\gamma_1$ are included, there is a small overlap
at zero energy
described by Eq. (\ref{eq:overlap}). In the full parameter model
(d), there is a similar amount of band overlap while the exact magnitude
of momentum at the crossing point
$vp \sim \Delta_1$ varies with angle
$\theta$ in a trigonal manner, and there is a tiny gap at those
crossing points. In (d), the self-consistent calculations yield
tiny $\Delta_2 < 0.01\gamma_1$ due to
non-zero $\gamma_2$, $\gamma_5$ and $\delta$.
%meaning that there are different densities on the
%layers $n_2 = - (n_1 + n_3)$ and $n_1 = n_3$ for
%$\Delta^{\rm (ext)} = 0$ and $n_{\rm tot} = 0$.

For each band structure we estimate the density of states (DOS) and conductivity
using the self-consistent Born approximation \cite{Shon,Kosh_bilayer}.
We assume that the scatterers are on-site potentials
localized on each layer, which is modeled by
$V(\Vec{r}) = \sum_{m=1,2,3} \sum_i u_i \delta(\Vec{r}-\Vec{r}_i) \hat{P}^{(m)}$
where $u_i$ and $\Vec{r}_i = (x_i, y_i)$ are the amplitude and
the two-dimensional position of the $i$-th scatterer, respectively,
and $\hat{P}^{(m)}$ is the projection operator onto the $m$-th layer.
We neglect inter-valley scattering between $K_\pm$.
The disorder strength is characterized by
$W = n_{\rm imp} u^2/(4\pi \hbar^2 v^2)$ \cite{Shon}
where $n_{\rm imp}$ is the total number of scatterers
over all the layers, and $u = \langle u_i^2 \rangle$.
For the model with only $\gamma_0$ and $\gamma_1$,
the energy scale for the level broadening at zero energy is given by
$\Gamma \sim (\pi/\sqrt{2}) W\gamma_1$.
Following Ref. \onlinecite{Shon},
we compute the self-energy and the vertex corrections for the
velocity operators, and calculate the conductivity using the Kubo formula.

%%%%%%%%%%%%%%%%%%%%%%%%%%%%%%%%%%%%%%%%%%%%%%%%%%%%%%%%%%%%%%%%%%%%%%%%%%%%%%
\begin{figure}[t]
%\centerline{\epsfxsize=0.95\hsize \epsffile{fig_cond_dos.eps}}
%\centerline{\epsfxsize=0.9\hsize \epsffile{fig3.eps}} 
\centerline{\epsfxsize=0.9\hsize \epsffile{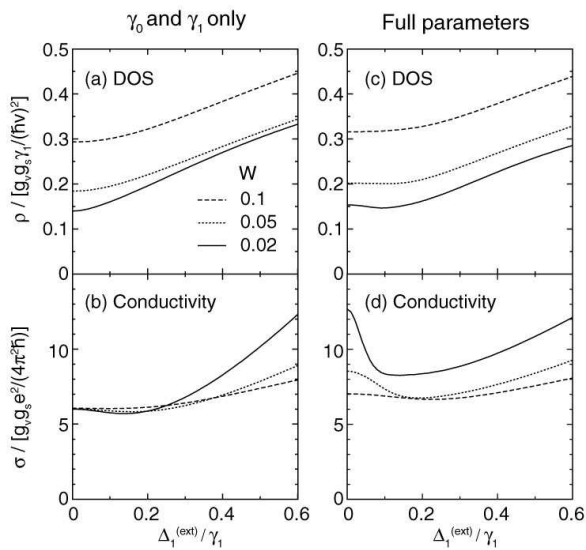}} 
\caption{
(a),(c) Density of states and (b),(d) conductivity at $n_{\rm tot}
= 0$ as functions of $\Delta_1^{\rm (ext)}$ for the model
including only $\gamma_0$ and $\gamma_1$ (left) and the full
parameter model (right). Solid, dotted, and dashed lines are for
progressively larger disorder strengths.} \label{fig:3}
\end{figure}
%%%%%%%%%%%%%%%%%%%%%%%%%%%%%%%%%%%%%%%%%%%%%%%%%%%%%%%%%%%%%%%%%%%%%%%%%%%%%%

Figures~\ref{fig:3}(a) and ~\ref{fig:3}(c) show the DOS
as a function of $\Delta_1^{\rm (ext)}$
at $n_{\rm tot} = 0$, for several values of the disorder strength $W$.
The left and right panels correspond to
the simple model with $\gamma_0$ and $\gamma_1$,
and the full parameter model discussed previously,
although the behavior in each case is
similar, DOS increases with $\Delta_1^{\rm (ext)}$.
Figure~\ref{fig:3}(b),(d) are plots of the conductivity
corresponding to (a),(c) respectively. The general trend
is for an {\it increase} of conductivity as $\Delta_1^{\rm (ext)}$
increases, except for the vicinity of $\Delta_1^{\rm (ext)}=0$
in panel (d).
This may be roughly understood by considering the
relation $\sigma = e^2 \rho_F v_F^2 \tau /2$ with velocity $v_F$,
DOS $\rho_F$ and relaxation time $\tau$ at the Fermi energy.
When we assume that all the states on the Fermi energy are
equally mixed by disorder,
we have $\tau \propto \rho_F^{-1}$, suggesting
that the conductivity is determined by $v_F^2$.
%In monolayers \cite{Shon} and bilayers \cite{Kosh_bilayer},
%assuming that states at the Fermi energy are fully mixed
%by disorder, $\tau \propto \rho_F^{-1}$ \cite{Shon},
%suggesting that the conductivity is determined by $v_F^2$.
The dispersion, Eq.~(\ref{eq:overlap}), approximates, in regions far
from the origin $vp \gg |\Delta_1|$, to $\epsilon \approx \pm
(v^2p^2-3\Delta_1^2/2)/(\sqrt{2}\gamma_1)$, showing that the
electron and hole bands are pushed towards zero energy by the
introduction of $\Delta_1$. This leads to an increase in the
expectation value of the band velocity in a disorder-broadened
energy window near $\epsilon = 0$, and thus the conductivity at
the charge neutral point is enhanced.

In the simple $\gamma_0$-$\gamma_1$ model, Fig.~\ref{fig:3}(b),
the conductivity at $\Delta_1^{\rm (ext)} = 0$ takes a universal
value $\sigma = 3g_vg_s e^2/(2\pi^2\hbar)$ independently of $W$.
This is because the Fermi energy coincides with the degeneracy
point of the monolayer and bilayer bands, and the value is indeed
equal to the summation of the minimum conductivity of monolayer
graphene \cite{Shon} and that of bilayer graphene
\cite{Kosh_bilayer} estimated in the self-consistent Born approximation. 
In Fig.~\ref{fig:3}(d), for
the full parameter model, the conductivity is largely enhanced
around $\Delta_1^{\rm (ext)} =0$, because, as observed in
Fig.~\ref{fig:2}(c), the Fermi energy crosses the off-center part
of the monolayer-like band
%owing to the relative shift of the monolayer and bilayer bands,
making a large contribution to the typical band velocity. The
conductivity drops sharply as $\Delta_1^{\rm (ext)}$ grows from
zero as the monolayer band is gapped away. When $\Delta_1^{\rm
(ext)}$ is increased further, the conductivity grows similarly to,
but a little more slowly than, Fig.~\ref{fig:3}(b), because of the
tiny gap at the band crossing point observed in
Fig.~\ref{fig:2}(d). The differences between Fig.~\ref{fig:3}(b)
and (d) become smaller for larger $W$, as
disorder-broadening %within a sufficiently large energy window
masks details dependent on the precise values of band parameters.

%%%

To conclude, we have shown that the breaking of mirror reflection
symmetry by interlayer asymmetry $\Delta_1$ in ABA-stacked
trilayer graphene causes hybridization of the linear and parabolic
bands, leaving just two bands in the vicinity of zero energy.
The band hybridization produces an increase in density of
states and typical band velocity with asymmetry $\Delta_1$,
leading to an increase in minimal conductivity
in qualitative agreement with
recent transport experiments \cite{crac08}. As demonstrated in
Fig. \ref{fig:1}(b), which compares the conductivities of trilayer
and bilayer graphene, the response of trilayers to gate-induced
asymmetry is in sharp contrast with bilayers, where the
conductivity is {\it suppressed} by a perpendicular electric field
owing to the opening of a gap between the electron and hole bands
\cite{oost08}.

The authors thank T. Ando, V.I.~Fal'ko, and H.~Schomerus for discussions, and
M.F.~Craciun, A.F.~Morpurgo, S.~Russo, and S.~Tarucha for discussions and
for sharing their experimental data prior to publication.
This project was funded by EPSRC-GB
First Grant No. EP/E063519/1, the Royal Society, and the Daiwa
Anglo-Japanese Foundation,
and by Grants-in-Aid
for Scientific Research from the Ministry of Education,
Culture, Sports, Science and Technology of Japan.


\begin{thebibliography}{99}

\bibitem{novo04} %K.S.~Novoselov {\em et al},
K.S.~Novoselov, A.K.~Geim, S.V.~Morozov,
D.~Jiang, Y.~Zhang, S.V.~Dubonos, I.V.~Grigorieva, A.A.~Firsov,
Science \textbf{306}, 666 (2004).

\bibitem{novo05}
%K.S.~Novoselov {\em et al},
K.S.~Novoselov, A.K.~Geim, S.V.~Morozov, D.~Jiang,
M.I.~Katsnelson, I.V.~Grigorieva, S.V.~Dubonos, and A.A.~Firsov,
Nature \textbf{438}, 197 (2005).

\bibitem{zhang05}
Y.B.~Zhang, Y.W.~Tan, H.L.~Stormer, P.~Kim, Nature \textbf{438}, 201 (2005).

\bibitem{novo06}
%K.~S.~Novoselov {\em et al},
K.~S.~Novoselov, E.~McCann, S.V.~Morozov, V.I.~Fal'ko,
M.I.~Katsnelson, U.~Zeitler, D.~Jiang, F.~Schedin, A.K.~Geim,
Nat. Phys. \textbf{2}, 177 (2006).

\bibitem{han07} M.Y.~Han, B.~Ozyilmaz, Y.~Zhang, and P.~Kim, Phys. Rev. Lett. \textbf{98},
206805 (2007).

\bibitem{miao07} % F.~Miao {\em et al}, 
F.~Miao, S.~Wijeratne, Y.~Zhang, U.~C.~Coskun, W.~Bao, and
C.~N.~Lau,
Science \textbf{317}, 1530 (2007).

\bibitem{stamp08} % C.~Stampfer {\em et al}, 
C.~Stampfer, J.~Guettinger, F.~Molitor, D.~Graf, T.~Ihn,
and K.~Ensslin,
Appl. Phys. Lett. \textbf{92}, 012102, (2008).

\bibitem{pono08} % L.A.~Ponomarenko {\em et al}, 
 L.~A.~Ponomarenko, F.~Schedin, M.~I.~Katsnelson, R.~Yang, E.~H.~Hill,
K.~S.~Novoselov, and A.~K.~Geim, 
Science \textbf{320}, 356 (2008).

\bibitem{mcc06a} E.~McCann and V.I.~Fal'ko, Phys. Rev. Lett. \textbf{96},
086805 (2006).

\bibitem{lu06} % C.L.~Lu {\em et al}, 
C.L.~Lu, C.P.~Chang, Y.C.~Huang, R.B.~Chen, and M.L.~Lin,
Phys. Rev. B \textbf{73}, 144427 (2006).

\bibitem{guinea06} 
F.~Guinea, A.H.~Castro~Neto, and N.M.R.~Peres, 
Phys. Rev. B \textbf{73}, 245426 (2006).

\bibitem{mcc06b} E. McCann, Phys. Rev. B \textbf{74}, 161403(R)
(2006).

\bibitem{min07} H.~Min, B.R.~Sahu, S.K.~Banerjee, and A.H.~MacDonald, 
Phys. Rev. B \textbf{75}, 155115 (2007).

\bibitem{castro07} %E.V.~Castro {\em et al},
E.V.~Castro, K.S.~Novoselov, S.V.~Morozov,
N.M.R.~Peres, J.M.B.~Lopes dos Santos, J.~Nilsson, F.~Guinea,
A.K.~Geim, A.H.~Castro Neto, 
Phys. Rev. Lett. \textbf{99}, 216802 (2007).

\bibitem{ohta06} %T.~Ohta {\em et al}, 
T.~Ohta, A.~Bostwick, T.~Seyller, K.~Horn, and E.~Rotenberg,
Science {\bf 313}, 951 (2006).

\bibitem{oost08} %J.B.~Oostinga {\em et al},
J.B.~Oostinga, %{\it et al.}, 
H.B.~Heersche, X.~Liu, A.F.~Morpurgo and L.M.K.~Vandersypen,
Nature Mater. {\bf 7}, 151 (2008).

\bibitem{ohta07} %T.~Ohta \textit{et al.}, 
T.~Ohta, A.~Bostwick, J.~L.~McChesney, T.~Seyller, K.~Horn, and E.~Rotenberg
Phys. Rev. Lett. \textbf{98}, 206802 (2007).

\bibitem{guett08} %J.~Guettinger {\em et al}, arXiv:0806.1384
J.~Guettinger, C.~Stampfer, F.~Molitor, D.~Graf, T.~Ihn,
and K.~Ensslin, %arXiv:0806.1384 
New. J. Phys. {\bf 10}, 125029 (2008).

\bibitem{crac08} %M.F.~Craciun {\em et al}, unpublished.
M.F. Craciun, S. Russo, M. Yamamoto, J.B. Oostinga, 
A.F. Morpurgo and S. Tarucha, 
to be published in Nature Nanotechnology.

\bibitem{latil06} S.~Latil and L.~Henrard, Phys. Rev. Lett. \textbf{97},
036803 (2006).

\bibitem{part06} B.~Partoens and F.M.~Peeters, Phys. Rev. B \textbf{74},
075404 (2006); \textbf{75}, 193402 (2007).

\bibitem{Kosh_mlg}
M.~Koshino and T.~Ando,
Phys. Rev. B {\bf 76}, 085425 (2007);
{\bf 77}, 115313 (2008).

\bibitem{aoki07} M.~Aoki and H.~Amawashi, Solid State Commun.
\textbf{142}, 123 (2007).

\bibitem{dressel02} M.S.~Dresselhaus and G.~Dresselhaus,
Adv. Phys. \textbf{51}, 1 (2002).

\bibitem{kpoints} Corners of the hexagonal Brillouin zone are located at
wave vector $\mathbf{K}_{\xi }=\xi ({\textstyle\frac{4}{3}}\pi a^{-1},0)$,
where $\xi =\pm 1$ and $a$ is the lattice constant.

\bibitem{guinea07} F.~Guinea, Phys. Rev. B \textbf{75}, 235433 (2007).

\bibitem{Shon}
N.H.~Shon and T.~Ando, J. Phys. Soc. Jpn. {\bf 67}, 2421 (1998);
Y.~Zheng and T.~Ando, Phys. Rev. B {\bf 65}, 245420 (2002).

\bibitem{Kosh_bilayer}
M.~Koshino and T.~Ando
Phys. Rev. B {\bf 73}, 245403 (2006).

\end{thebibliography}
\end{document}